# Prediction of $CO_2$ reduction reaction intermediates and products on transition metal-doped γ-GeSe monolayers: A combined DFT and machine learning approach

Xuxin Kang,[a] Wenjing Zhou,[a] Ziyuan Li,[a] Zhaoqin Chu,[b] Hanqin Yin,[c] Shan Gao,[a] Aijun Du,[c,*] Xiangmei Duan[a,*]

[a] School of Physical Science and Technology, Ningbo University, Ningbo, 315211, China.

[b] Zhejiang Key Laboratory of Data-Driven High-Safety Energy Materials and Applications, Advanced Interdisciplinary Sciences Research (AiR) Center, Ningbo Institute of Materials Technology and Engineering, Chinese Academy of Sciences, Ningbo 315201, China.

[c] School of Chemistry and Physics and Centre for Materials Science, Queensland University of Technology, Gardens Point Campus, Brisbane, QLD 4001, Australia

Email: aijun.du@qut.edu.au, duanxiangmei@nbu.edu.cn Tel: 13429322782

## Abstract

The electrocatalytic $CO_2$ reduction reaction ($CO_2RR$) is a complex multi-proton-electron transfer process that generates a vast network of reaction intermediates. Accurate prediction of free energy changes ($\Delta G$) of these intermediates and products is essential for evaluating catalytic performance. We combined density functional theory (DFT) and machine learning (ML) to screen 25 single-atom catalysts (SACs) on defective γ-GeSe monolayers for $CO_2$ reduction to methanol, methane, and formic acid. Among nine ML models evaluated with 14 intrinsic and DFT-based features, the XGBoost performed best ($R^2$ = 0.92 and MAE = 0.24 eV), aligning closely with DFT calculations and identifying Ni, Ru, and Rh@GeSe as prospective catalysts. Feature importance analysis in free energy and product predictions highlighted the significance of $CO_2$ activation with ∠O−C−O and $IP^{C-O1}$ as the key attributes. Furthermore, by incorporating non-DFT-based features, rapid predictions became possible, and the XGBoost model retained its predictive performance with $R^2$ = 0.89 and MAE = 0.29 eV. This accuracy was further validated using Ir@GeSe. Our work highlights effective SACs for $CO_2RR$, and provides valuable insights for efficient catalyst design.

# 1. Introduction

The rapid depletion of fossil fuels has driven an unprecedented surge in greenhouse gas emissions, exacerbating concerns regarding sea-levels rising and global warming. Elevated $CO_2$ levels, the primary contributor to the greenhouse effect, pose a grave threat to human and ecological health. To address this, sustainable strategies for $CO_2$ capture and conversion are urgently needed.[1-6] Electrochemical $CO_2$ reduction reaction ($CO_2$RR) to valuable chemicals and fuels stands out as a green and promising approach.[7-9] However, $CO_2$RR involves a complex multi-proton-electron transfer (MPET) process, yielding many different products such as CO, formic acid (HCOOH), methanol ($CH_3OH$), and methane ($CH_4$). Despite advances in diverse materials, challenges in catalytic selectivity and activity persists due to the intricate reaction pathways and the high stability of $CO_2$ molecule requiring significant activation energy. Additionally, competitive hydrogen evolution reactions (HER) can dominate, and thus impede the $CO_2$RR at higher voltages. Therefore, the development of environmentally friendly catalysts with enhanced efficiency for $CO_2$RR is a critical research priority.[10-15]

Since their inception in 1978, metal-support interactions (MSIs) have garnered considerable attention for modulating the electronic and catalytic properties of catalysts.[16] Recently, MSIs have obtained renewed interest in single atom catalysts (SACs), where isolated metal atoms are stabilized on pristine or defective substrates.[17-21] The strength of MSIs, influenced by the substrate's charge transfer and coordination environment, alters the electronic states of transition metal (TM) atoms, significantly affecting catalytic performance. For instance, Xun et al. reported a limiting potential ($U_L$) of −1.24 V for $CO_2$ reduction to $CH_4$ on a Ni supported $MoSi_2N_4$ monolayer,[22] while Kou et al. achieved lower potentials of −0.36 V and −0.43 V for Ni and Rh supported $In_2Se_3$ monolayers.[23] Lei et al. demonstrated exceptional HCOOH selectivity with a $U_L$ of −0.31 V for a single Ni atom anchored on defective phosphorene monolayer, compared to a higher $U_L$ of −0.98 V for producing $CH_4$,[24] and the Pan group reported $U_L$ values of −0.29 V, −0.75 V, and −0.74 V for producing HCOOH, $CH_4$, and $CH_3OH$ on Ni, Ru, and Rh anchored $Ti_2CO_2$.[25] These findings underscore the importance of substrate selection in the design of SACs. However, the limited understanding of the $CO_2$RR mechanism and factors governing catalytic activity complicates the development of efficient catalysts. Identifying accessible descriptors to accurately predict the free energy changes of intermediates and products is essential for high-throughput screening of catalysts with improved catalytic activity and selectivity.

To date, research on group IV-VI compounds has primarily focused on the α-phase, akin to black phosphorus, known for its strong in-plane ferroelectric polarization. However, Zou et al.

theoretically predicted a novel γ-phase structure with a four-atomic-layer thickness and unique band dispersion,[26] later synthesized by Kim et al as a 2D γ-phase GeSe monolayer via chemical vapor deposition.[27] Unlike the conventional α- and β-phases, γ-GeSe exhibits a narrower bandgap and exceptional electronic conductivity, rivaling even most semi-metallic layered materials and surpassing graphite.[27] Subsequent studies have explored its electrical, optical, thermal, and thermoelectric properties in pristine, strained, and defective γ-GeSe monolayers.[28-32] Cai et al. further proposed the γ-GeSe monolayer as a high-capacity anode for lithium-ion batteries due to its low diffusion barrier and excellent conductivity.[33] Inspired by these properties, we hypothesize that the defective γ-GeSe monolayer is a promising substrate for supporting SACs towards efficient $CO_2RR$.

In this work, we systematically evaluated 25 TMs from the 3d, 4d, and 5d series anchored on defective γ-GeSe monolayers (denoted as TM@GeSe) for $CO_2$ reduction to C1 products, using DFT and ML models. Through rigorous evaluation of their stability, $CO_2$ activation, selectivity, and activity, Ni, Ru, and Rh@GeSe emerged as top candidate catalysts for $CO_2RR$. We trained ML models using two schemes: one with 14 features, including intrinsic and DFT-based (IF+DFT), the other with only 11 intrinsic features (IF), testing nine algorithms on 196 free energy values, and found XGBoost model exhibited exceptional predictive performance for both schemes. Additionally, we defined the producing probability ($P$) and employed the XGBoost model to predict products. Our work offers a robust framework in catalyst design through rapid, and accurate predictions.

## 2. Calculation methods

The spin-polarized DFT calculations were performed using the Vienna *Ab initio* Simulation Package (VASP) with the projector-augmented wave (PAW) method.[34-35] The Perdew-Burke-Ernzerhof (PBE) exchange-correlation functional within the generalized gradient approximation (GGA) was employed.[36] A cutoff energy of 520 eV was applied, and the convergence criteria were set as $10^{-5}$ eV for energy and 0.01 eV/Å for force. The first Brillouin zone was sampled using a 3×3×1 Monkhorst−pack k–point grid for structure relaxation and a 6×6×1 grid for density of states calculation. To avoid periodic interaction effects, a vacuum space of approximately 15 Å along the z-direction was incorporated. The van der Waals D3 (DFT+D3) correction was applied for accurate modeling of small molecule adsorption.[37] Charge density difference and Bader charge analysis were implemented to illustrate the charge transfer. The strength of the chemical bond between $CO_2$ and TMs was evaluated using the projected crystal orbital Hamilton population (pCOHP) obtained by the LOBSTER code.[38] To evaluate the diffusion ease of TMs between vacancies, the diffusion barriers of TMs were determined based on the climbing image-nudged elastic band (CI-NEB) method.[39] Phonon spectrum of γ-GeSe monolayer was calculated using the PHONOPY program with a 4×4×1 supercell. *Ab initio* molecular dynamics simulations (AIMD) were conducted at 500 K with a total time step of 10 ps to assess the thermodynamic stability of SACs.

The Δ$G$ of every fundamental step of $CO_2$RR was obtained based on the computational hydrogen electrode (CHE) model proposed by Nørskov et al.[40] Detailed calculation process can be found in the Supporting Information. The implicit solvation models in VASPsol were constructed to examine the effect of solvation on catalytic activity, employing a dielectric constant of 80 for water.[41]

All ML algorithms, including Artificial neural networks (ANNs),[42] Gradient boosting regression (GBR),[43] Extreme gradient boosting regression (XGBoost),[44] Least absolute shrinkage and selection operator (LASSO),[45] Support vector regression (SVR),[46] Decision-trees (DT),[47] Extra-trees (ET),[48] K-nearest neighbor (KNN),[49] Random forest regression (RFR),[50] were executed using the open-source Scikit-learn library within a Python3 environment.[51] The datasets, collected through DFT calculations, were randomly split into training and test sets with a ratio of 80:20. To assess the performance of these models, we selected three key metrics: coefficient of determination values ($R^2$), mean absolute error (MAE), and root-mean-square error (RMSE), as outlined in the following equation:

$$R^2 = 1 - \frac{\sum_{i=1}^{n}(\hat{y}_i - y_i)^2}{\sum_{i=1}^{n}(\overline{y}_i - y_i)^2} \quad (1)$$

$$\text{MAE} = \frac{1}{n}\sum_{i=1}^{n}|y_i - \hat{y}_i| \quad (2)$$

$$\text{RMSE} = \sqrt{\frac{1}{n}\sum_{i=1}^{n}|y_i - y_i|^2} \quad (3)$$

where $\hat{y}_i$ represents the *i*-th calculated value by DFT, $y_i$ means the *i*-th predicted value by ML model, and $\overline{y}_i$ denotes the average value of DFT calculation results.

## 3. Results and Discussion

We selected the 2D γ-phase GeSe monolayer as a potential substrate for SACs on three compelling factors: Firstly, the γ-phase is the ground state, exhibiting a total energy lower by 0.31 eV and 0.51 eV compared to the α- and β-phases, respectively, aligning with previous reports.[26] Secondly, the remarkable electronic conductivity has been experimentally demonstrated in synthesized γ-phase GeSe monolayers,[27] which is confirmed in the calculated energy band diagram in Figure S1(a). Thirdly, its exceptional dynamical stability is evidenced by the absence of imaginary frequencies in the phonon spectrum, as depicted in Figure S1(b).

As illustrated in Figure S2(a), the optimized γ-GeSe monolayer features a distinctive sandwich structure composed of two interconnected single-layer honeycomb lattices. The calculated lattice constants are $a = b = 3.79$ Å, with a thickness of 4.69 Å, and the bond lengths of Ge−Se and Ge−Ge are 2.58 Å and 2.95 Å, respectively, which are consistent with previous studies.[30, 32]

For efficient $CO_2$ reduction, effective capture and activation of $CO_2$ molecules on the catalyst surface are crucial. However, as depicted in Figure S3, when the $CO_2$ molecule is adsorbed on either Se or Ge atoms, it retains its linear structure, suggesting that the pristine 2D γ-GeSe monolayer may not be an effective catalyst for $CO_2RR$.

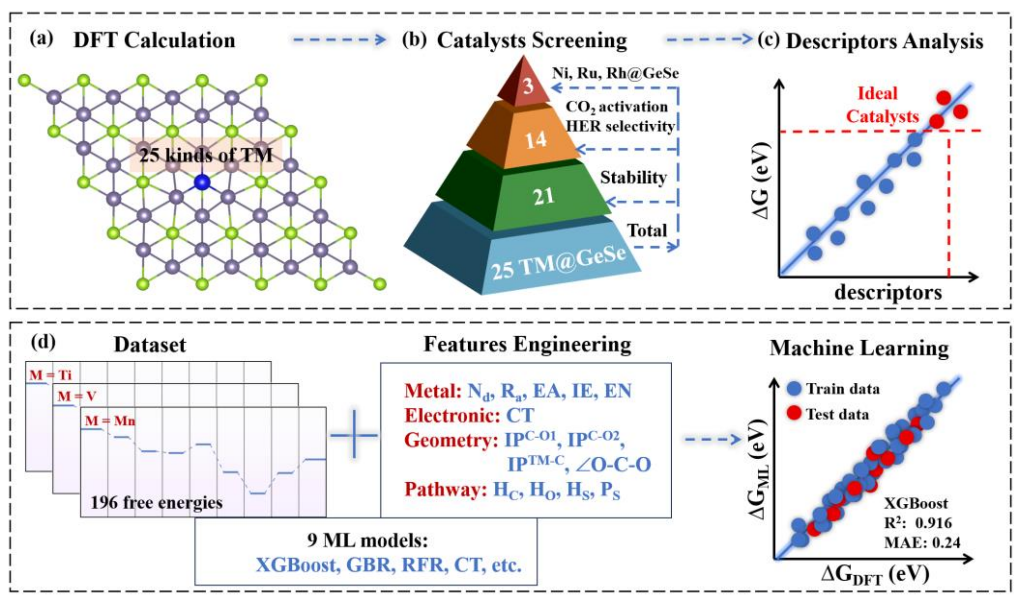

**Figure 1.** The flowchart used in the work. (a) The construction of 25 different single-atom catalysts. (b) Screening process of promising catalysts. (c) Analysis of key descriptors affecting $\Delta G$ values of intermediates. (d) Flow of the machine learning approach.

### 3.1 Structure and Stability of SACs

We consider anchoring TMs into a single Se vacancy ($V_{Se}$) on the 4×4×1 GeSe monolayer (Figure 1a), where TMs supply additional electrons to break the strong $sp$ hybridization. The formation energy ($E_f$) of $V_{Se}$ was calculated to be 1.3 eV, comparable to those of $V_S$-$MoS_2$, $V_{Se}$-$MoSe_2$, and $V_{Se}$-InSe,[52-54] and significantly lower than that of $V_C$ in graphene and $BC_3$.[55-56] This suggests the feasibility of $V_{Se}$ formation.

A total of 25 SACs were constructed, and their stability, activity, and selectivity were rigorously evaluated, leading to the identification of the most promising candidates (Figure 1b). The interactions between TMs and substrate were evaluated by calculating the binding energy ($E_b$). As shown in Table S1, all SACs exhibited considerably negative $E_b$ compared to previously proposed SACs,[57-58] indicating a high thermodynamic stability of TM@GeSe. To further assess the tendency of metal atoms to aggregate into clusters, the ratio between $E_b$ and cohesive energy ($E_c$) was calculated. The relatively high $E_b/E_c$ values, ranging from 0.6 to 1.19, indicated that TMs preferred to remain isolated on the substrate.[56, 59]

To substantiate the practical feasibility of TM@GeSe, Ni, Ru, and Rh@GeSe (Figure S4) were exemplified, and the diffusion barriers of TMs across adjacent vacancies were computed. The corresponding diffusion barriers were as high as 2.01 eV, 1.80 eV, and 4.20 eV, respectively, indicating effective inhibition of TM aggregation.[23] As displayed in Figure S5, the energy and temperature simulated by AIMD at 500 K oscillated only slightly around the equilibrium positions, and the structures did not manifest significant deformation or fracture after running 10 ps, revealing high thermodynamic stability and the potential for experimental synthesis of TM@GeSe.[20, 23, 60] Furthermore, the electrochemical stability was evaluated by computing the dissolution potential ($U_{diss}$), where positive values indicate that the metal atoms will not detach from the substrate and dissolve into the electrolyte under acidic conditions.[61-62] As evident from Table S1, most SACs exhibited superior electrochemical stability, except for Sc, Y, Zr, and Hf@GeSe, which were not further elaborated upon.

## 3.2 Adsorption and Activation of $CO_2$ Molecules

The $CO_2$ molecules can bind to TMs in different adsorption modes, including physisorption, end-on, side-on1, and side-on2, as displayed in Figure 2(a). The optimized structure and parameters of $CO_2$ adsorption, such as TM−C and TM−O1 bond lengths, C−O1 and C−O2 bond lengths, and ∠O−C−O bond angles, are summarized in Table S2 and Figure S6.

Weak physisorption can occur on Pt and Au, evidenced by the large distance (more than 3 Å) between the TMs and $CO_2$ molecules. The C−O bond lengths and ∠O−C−O bond angles are close to 1.17 Å and 180° of free $CO_2$ gas molecules, respectively. As depicted in Figure 2(b), electron transfer between Pt@GeSe and $CO_2$ is negligible.

$CO_2$ can also form weak adsorption on TM@GeSe in the end-on mode, where an oxygen atom of $CO_2$ attaches directly to TMs while maintaining a nearly linear shape. Cr, Cu, Pd, and Ag prefer this mode, with TM−O1 bonds ranging from 2.20 Å to 2.84 Å. The C−O1 bonds stretch slightly to 1.18 Å ~ 1.19 Å, while C−O2 bond lengths remain almost the same. As illustrated in Figure 2(b) for Cr@GeSe, charge transfer primarily happens between the C and O1 atoms, not from TM to O1, denoting local activation of the $CO_2$ molecule.

Strong $CO_2$ chemisorption occurs on TM@GeSe (where TM = Ni, Ru, Rh, and Ir) in the side-on1 mode, where only the carbon atom binds directly to the TMs. This results in two equal C−O bond lengths between 1.20 Å and 1.22 Å, and ∠O−C−O bond angles between 149.23° and 157.39°. For TM like Ti, V, Mn, Fe, Co, Nb, Mo, Ta, W, Re, and Os, $CO_2$ adsorption happens in the side-on2 mode. Both the carbon and oxygen atoms of $CO_2$ bind simultaneously to TMs, causing bond angles between 128.20° and 147.46°, longer C−O1 bonds (above 1.26 Å), and slightly longer C−O2 bonds (about 1.21 Å). Charge density difference plots of Ni and Mo@GeSe show more electron transfer, indicating $CO_2$ molecules are effectively activated in both side-on modes. So, except for Pt, Au@GeSe (physisorption) and Cr, Cu, Pd, Ag@GeSe (local activation), the mechanism of the remaining 15 SACs was investigated further.

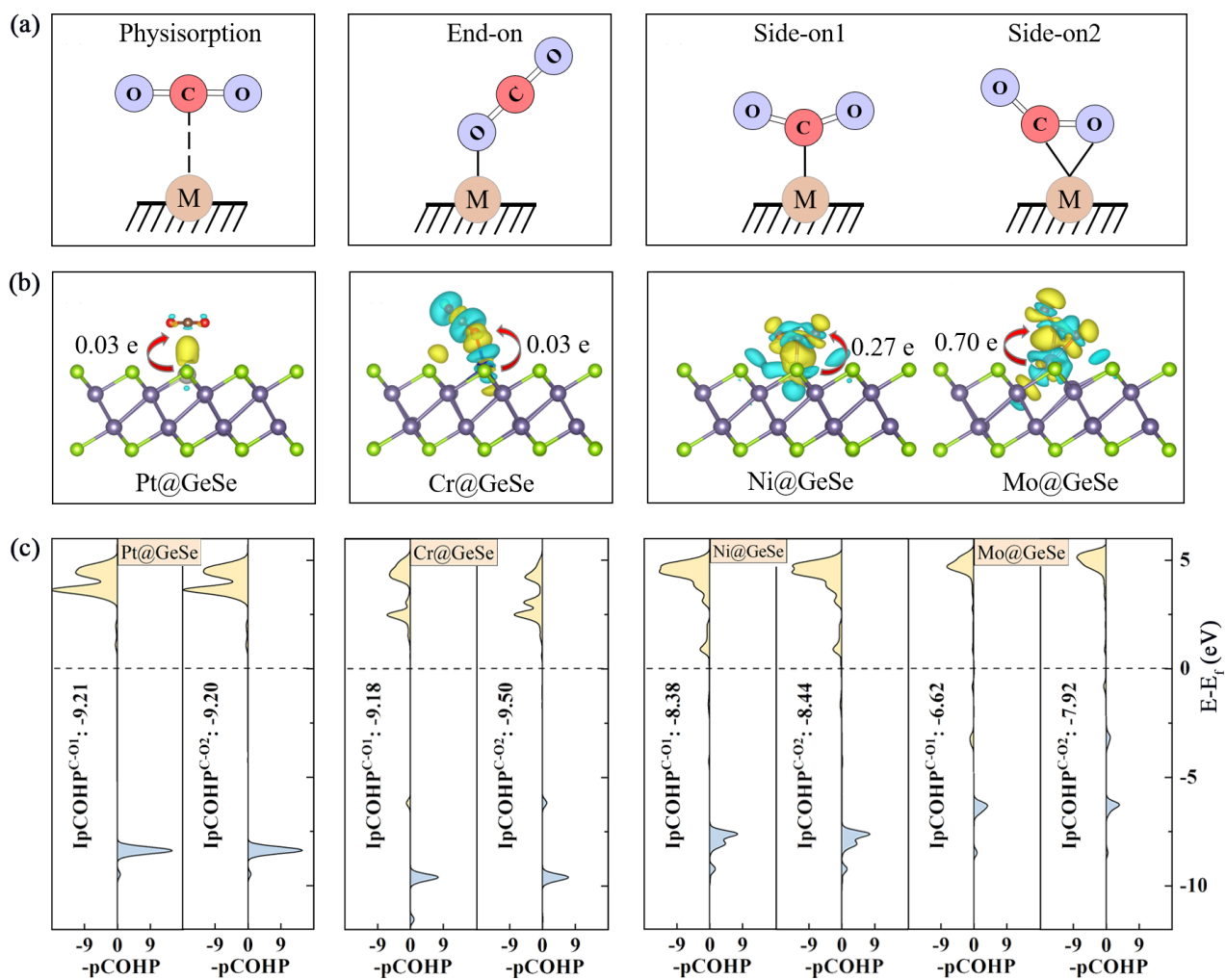

**Figure 2.** (a) Four adsorption configurations for $CO_2$ on a catalyst surface. (b) Charge density difference diagrams for $CO_2$ adsorption on Pt@GeSe, Cr@GeSe, Ni@GeSe, and Mo@GeSe. Yellow and cray show charge accumulation and depletion, respectively (isosurface value: 0.005 e Å$^{-3}$). (c) Projected crystal orbital Hamilton population (pCOHP) analysis of C−O bonds after $CO_2$ adsorption.

### 3.3 Electronic Structure of $CO_2$ Activation

We delved into the electronic-level mechanism of $CO_2$ adsorption and activation by analyzing the projected density of states (PDOS) and projected crystal orbital Hamilton population (pCOHP) of the adsorbed systems. The PDOS of examples like Pt, Cr, Ni, Mo, Ru, and Rh@GeSe, are presented in Figure S7. The hybridization between the TM-$d$ and $CO_2$-$2p$ orbitals is prominent for Ni, Mo, Ru, and Rh@GeSe, indicating a strong interaction between these SACs and $CO_2$ in side-on modes. Notably, in these four systems, the antibonding states of the C−O bonds are near or below the Fermi energy level, meaning the C−O bonds are less stable and the $CO_2$ molecule is effectively activated, as shown in Ni, Mo@GeSe in Figure 2(c) and Ru, Rh@GeSe in Figure S7.

The integral pCOHP (IpCOHP) values for interactions between the TM and its coordinated carbon atoms (denoted as $IP^{TM-C}$) and oxygen atoms ($IP^{TM-O1}$) were calculated. As shown in Table S3, the $IP^{TM-C}$ and $IP^{TM-O1}$ values for both side-on modes are more negative than those for end-on and physisorption modes, indicating the interactions are stronger. Also, the IpCOHP values between carbon and both O1 ($IP^{C-O1}$) and O2 ($IP^{C-O2}$) were computed to assess the degree of activation. In the end-on and physisorption modes, $IP^{C-O1}$ and $IP^{C-O2}$ values are close to −9.21 eV, which is the value for free $CO_2$ molecules. But for both side-on modes, these values are much higher, signifying effective $CO_2$ activation.

### 3.4 Screening of Promising Catalysts

During $CO_2RR$, HER is a major competing side reaction. In SACs, hydrogen atoms on the metal center can poison the active site, reducing the $CO_2RR$ efficiency. The first step in $CO_2$ hydrogenation forms either *OCHO or *OCOH species. Usually, a lower $\Delta G$ value means better stability and selectivity.[63-64] To screen catalysts with high selectivity, we initially compared the $\Delta G$ values of *OCOH, *OCHO, and *H species. As depicted in Figure 3(a), among 15 screened SACs, the first protonation step favors forming *OCHO intermediates because of their lower $\Delta G$. So, only *OCHO intermediate was considered in subsequent steps. Furthermore, except for Ir@GeSe, the other 14 TM@GeSe catalysts exhibited superior selectivity for $CO_2RR$ over HER. This highlights the potential of using TM@GeSe as cathodes for high-efficiency $CO_2RR$.

We then calculated the $\Delta G$ of all intermediates for 14 SACs and summarized the most favorable reaction pathway, the potential determining step (PDS), $U_{Ls}$, and final products in Figure 3(b). The free energy diagrams are displayed in Figure 3, S8, and S9. Notably, Ni, Ru, and Rh@GeSe stood out as excellent catalysts with low $U_{Ls}$, similar to previously reported values.[23-24, 65-66] Specifically, Ni and Rh@GeSe catalyze $CO_2$ reduction to HCOOH with $U_{Ls}$ of −0.37 V and −0.41 V, respectively, while Ru@GeSe favors $CH_4$ production with a $U_L$ of −0.39 V. For more details on the hydrogenation process, see the Supporting Information.

Since $CO_2RR$ happens in water, we studied the effect of solvent on TM@GeSe catalyst using implicit solvation models in VASPsol, with a dielectric constant of 80 for water. The results for Ni, Rh, and Ru@GeSe are presented in Fig. S11. The solvation had little impact on the free energy of most intermediates, except for *OCHOH. But the $U_L$ decreased when considering the solvation effect. Still, this did not change the overall catalytic performance of $CO_2RR$ on TM@GeSe much.

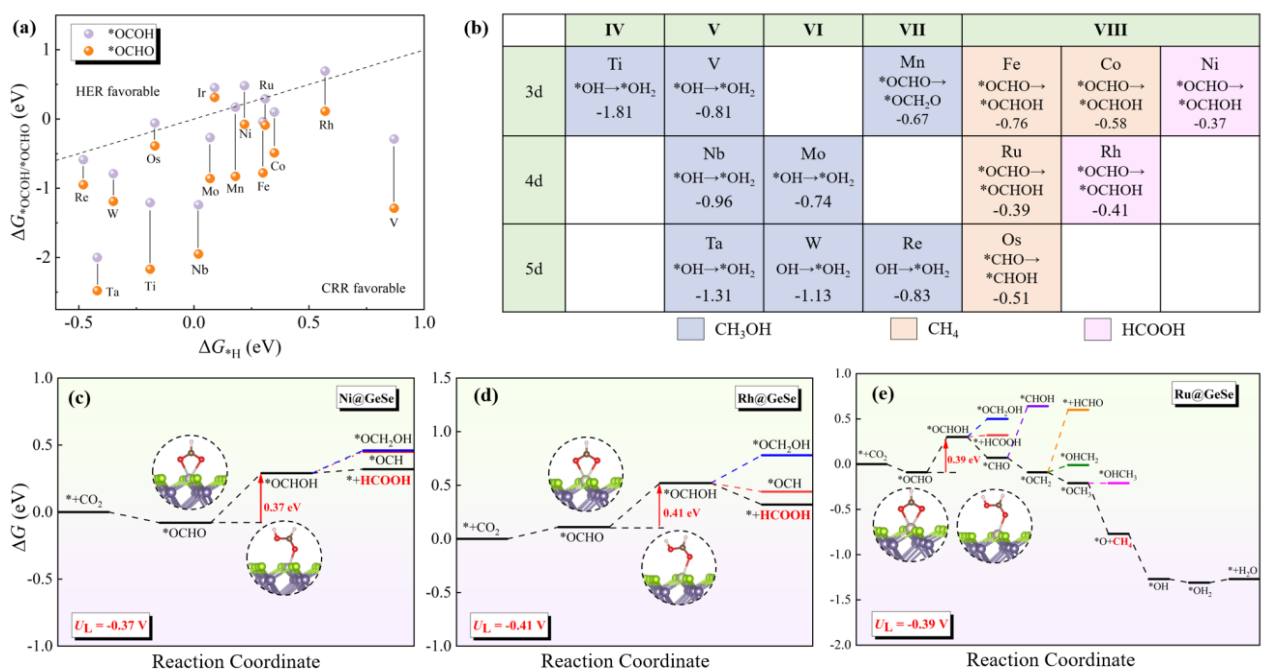

**Figure 3.** (a) Selectivity analysis for CO$_2$RR depicted as the $\Delta G_{*OCOH}$ or $\Delta G_{*OCHO}$ compared to the $\Delta G_{*H}$ of HER. (b) Summary of the PDS, $U_{L}$s, and final products for 14 SACs. Free energy diagrams for CO$_2$RR on (c) Ni@GeSe, (d) Rh@GeSe, and (e) Ru@GeSe, showing intermediate structures at the initial and final states of PDS.

### 3.5 Descriptors Analysis in CO$_2$RR Process

Because the CO$_2$RR pathway is complex, calculating all possible reaction intermediates requires massive computational resources, especially for high-throughput screening of potential candidates. Prior studies have highlighted the importance of understanding the intrinsic activity of SACs by using the right descriptors.[67] To elucidate the structure-activity relationships, we employed specific descriptors and ML models to simplify the mechanism and facilitate the rational design of CO$_2$RR electrocatalysts, as shown in Figure 1(c) and 1(d).

*OCHO and *OH intermediates are crucial in CO$_2$RR, as evidenced by their strong linear correlation with other intermediates' $\Delta G$ values (Figure S12). The efficiency of product desorption depends on the adsorption strength of these intermediates. If the adsorption is too strong, it can block later reactions, increasing $U_L$. There is a positive correlation between each intermediate's $\Delta G$ and the $U_L$. For example, Ni, Ru, and Rh@GeSe catalysts, denoted by blue spheres, have $U_L$ below −0.5 V, suggesting weaker adsorption of intermediates lead to better catalytic performance.

The 2D volcano plots in Figure 4(a-f) illustrate $U_L$ for HCOOH, CH$_4$, and CH$_3$OH from CO$_2$RR as a function of $\Delta G_{*OCHO}/\Delta G_{*OH}$. Ni, Ru, and Rh@GeSe are at the top of the volcano plots for HCOOH and CH$_4$, indicating weaker adsorption of *OCHO and *OH favors these products. The data points along with the blue solid line suggest that for most TM@GeSe, the PDS in making HCOOH and CH$_4$ is the step from *OCHO to *OCHOH, while for Ti, V, and Re@GeSe, the PDS for CH$_4$ is the step from *OH to *OH$_2$. For CH$_3$OH production, moderate adsorption of *OCHO and *OH is preferred. Specifically, the PDS for pre-TM-doped GeSe is the step from *OH to *OH$_2$, while for post-TM-doped GeSe, it is the step from $^*$OCHO to $^*$OCH$_2$O. Remarkably, Mn, Mo, V, and Re@GeSe are at the volcano peak for CH$_3$OH production. The lowest solid lines in Figure 4(a-f) represent the basic reaction steps that limit CO$_2$RR activity, explaining why the second or last hydrogenation step is often the PDS in previous studies.[11,25] A 3D contour map in Figure 4(g) shows $U_L$ as a function of $\Delta G_{*OCHO}$ and $\Delta G_{*OH}$. The dark red triangular area in the center signifies the lowest $U_L$ values, representing the best catalytic activity. The selected Ni, Ru, and Rh@GeSe catalysts are close to this ideal area.

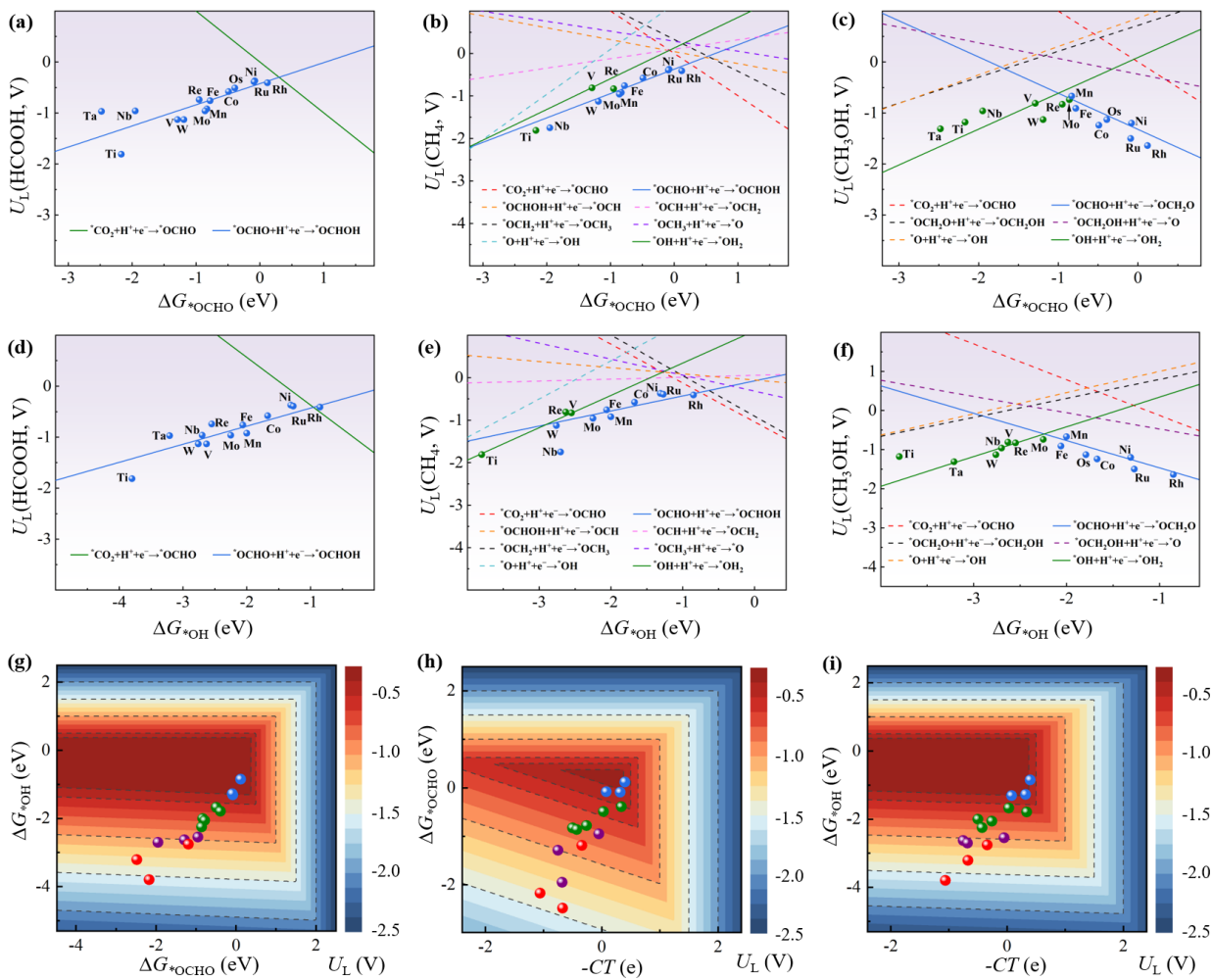

**Figure 4.** 2D volcano plots of $U_L$ for CO$_2$RR toward (a) HCOOH, (b) CH$_4$, and (c) CH$_3$OH as a function of $\Delta G_{*OCHO}$. (d-f) 2D volcano plots of $U_L$ versus $\Delta G_{*OH}$. Colored lines depict key reaction steps. (g) 3D contour plots of $U_L$ as a function of $\Delta G_{*OCHO}$ and $\Delta G_{*OH}$. (h-i) $U_L$ versus charge transfer (-$CT$), $\Delta G_{*OCHO}$ and $\Delta G_{*OH}$, respectively. Colored spheres show $U_L$ ranges.

The number of $d$ electrons (N$_d$) in TMs is strongly related to the $\Delta G$ values of intermediates, $U_L$s, and products (Figure 3b and S13). Early TMs (groups IV B – VII B) with lower N$_d$ favor CH$_3$OH production, while later TMs (group VIII B) with higher N$_d$ favor CH$_4$ and HCOOH formation. Also, TMs with fewer N$_d$ bind more strongly to intermediates, leading to higher $U_L$ (Figure S13).

Steps in CO$_2$RR that increase entropy can be described using the number of hydrogen atoms on carbon ($H_C$) and oxygen ($H_O$) in reaction intermediates, as well as the number of hydrogen atoms in $CH_4$, $CH_3OH$, and $H_2O$ species ($H_S$) released during the process. For example, steps like *$OCH_2OH$ ($H_C$ = 2, $H_O$ = 1, $H_S$ = 0) → *O + $CH_3OH$ ($H_C$ = 0, $H_O$ = 0, $H_S$ = 4), *$OCH_3$ ($H_C$ = 3, $H_O$ = 0, $H_S$ = 2) → *O + $CH_4$ ($H_C$ = 0, $H_O$ = 0, $H_S$ = 6) show these changes.

$CO_2RR$ is a MPET process. As illustrated in Figure S2(c), making HCOOH, $CH_3OH$, and $CH_4$ products requires 2, 6, and 8 proton-electron pairs, respectively. The first ($P_S$ = 1), fourth ($P_S$ = 4), and seventh ($P_S$ = 7) steps typically have negative $\Delta G$, meaning they are thermodynamically favorable. Especially early TMs show strong interactions between oxygen and the metal in intermediates like *OCHO, *$OCH_2OH$, *OCH, *$OCH_3$, *O, *OH.

Charge transfer (CT) values can be used to measure the impact of MSIs on catalytic performance. As depicted in Figure 4(h-i) and S13, the $\Delta G$ of intermediates and $U_L$ are strongly related to CT values. Early TMs, particularly Ti, V, Mn, Nb, and Ta, have CT values above 0.5, signifying they lose a lot of electrons, leading to stronger interactions with intermediates. In contrast, late TMs like Co, Ni, Ru, Rh, and Os gain electrons from the substrate, making their $d$-shells almost full. This means fewer empty orbitals are available for electron acceptor-donor interactions, leading to weaker adsorption of intermediates and easier product release.

To quantify the whole reaction process, we used descriptors like IpCOHP values for TM−C ($IP^{TM-C}$), C−O1 ($IP^{C-O1}$), C−O2 ($IP^{C-O2}$), and the bond angles of O−C−O (∠O−C−O). As shown in Figure S13, ∠O−C−O and $IP^{C-O1}$ values are more strongly related to the $\Delta G$ of intermediates than $IP^{TM-C}$ and $IP^{C-O2}$. Along with parameters mentioned earlier, four intrinsic features of TMs ($R_a$, EA, IE, $\chi$) were also used to predict free energy changes and product selectivity.

### 3.6 ML predictions of free energy changes and products

A total of 14 descriptors were chosen for feature engineering in ML models, as listed in Figure 5(a). Intrinsic features (IF) of TMs, including $N_d$, $R_a$, EA, IE, and $\chi$, are listed in Table S4. Pathway-specific features ($H_C$, $H_O$, $H_S$, and $P_S$) come from the $CO_2RR$ process. CT, $IP^{C-O1}$, $IP^{C-O2}$, $IP^{TM-C}$, and ∠O−C−O were derived from DFT calculations. Using a dataset of 196 free energy values and 14 features (IF+DFT), we built nine common ML models: ANNs, GBR, XGBoost, LASSO, SVR, DT, ET, KNN, and RFR. $R^2$, MAE, and RMSE metrics were used to evaluate the performance of models (Eqs. 1-3). A 10-fold cross-validation grid search was performed to optimize model parameters.

Among the models, XGBoost, GBR, and RFR performed the best, with high $R^2$ values (> 0.85) and low MAE and RMSE values (< 0.40), as shown in Figure 5 and S14. The XGBoost model was the best, with $R^2$ = 0.92, MAE = 0.24 eV, and RMSE = 0.31 eV. It also had better stability, with a higher mean cross-validation score (0.77) compared to GBR (0.74) and RFR (0.66). Features importance analysis for the XGBoost model revealed that ∠O−C−O, $IP^{C-O1}$, CT, and $N_d$ were the most important features (Figure 5c and S15). Among these, the ∠O−C−O and IP$^{C-O1}$ contributed 29.7% and 26.2%, respectively, underscoring the significance of $CO_2$ activation in $CO_2$RR. This matches the descriptor analysis.

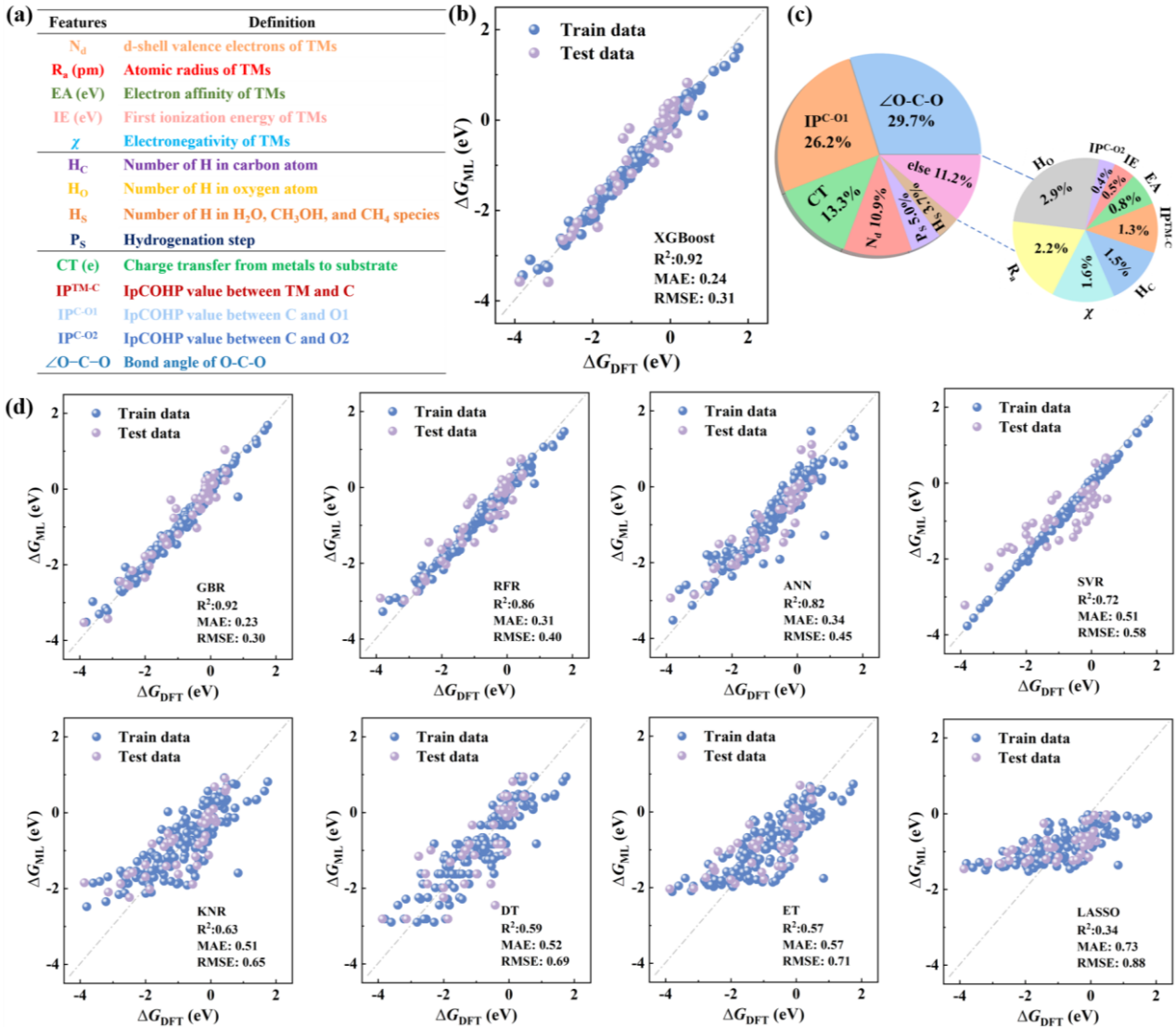

**Figure 5.** (a) ML model with 14 features including 9 intrinsic features and 5 DFT-based features. (b) Comparison of ΔG values between ML predictions and DFT calculations. (c) Features importance assessment for the XGBoost model. (d) Performance of the other 8 ML models in predicting ΔG.

The free energy diagrams in Figure S16 compare the XGBoost model predictions using IF+DFT features with DFT results, demonstrating good agreement for each ΔG. The PDS remained the same, and the $U_L$ showed a small error of about 0.15 eV. Also, the ML predictions identified Ni, Ru, and Rh@GeSe as promising catalysts.

To make faster predictions, we employed features that don't need DFT computations. As illustrated in Figure 6(a), DFT-based descriptors were replaced with non-DFT parameters. Specifically, the electronegativity ($\chi_{MC}$) and valance electrons ($VE_{MC}$) of the metal and its first coordinated atoms were incorporated to replace the CT descriptor, which describes MSIs based on equations 4-5.[68] Furthermore, the global electronegativity difference (Δ$\chi$) between catalysts and intermediates, and the average ionization energy difference (ΔIE) between metals and intermediates, were introduced to show the interaction strength of intermediates. In the end, 11 intrinsic features were chosen for the ML models.

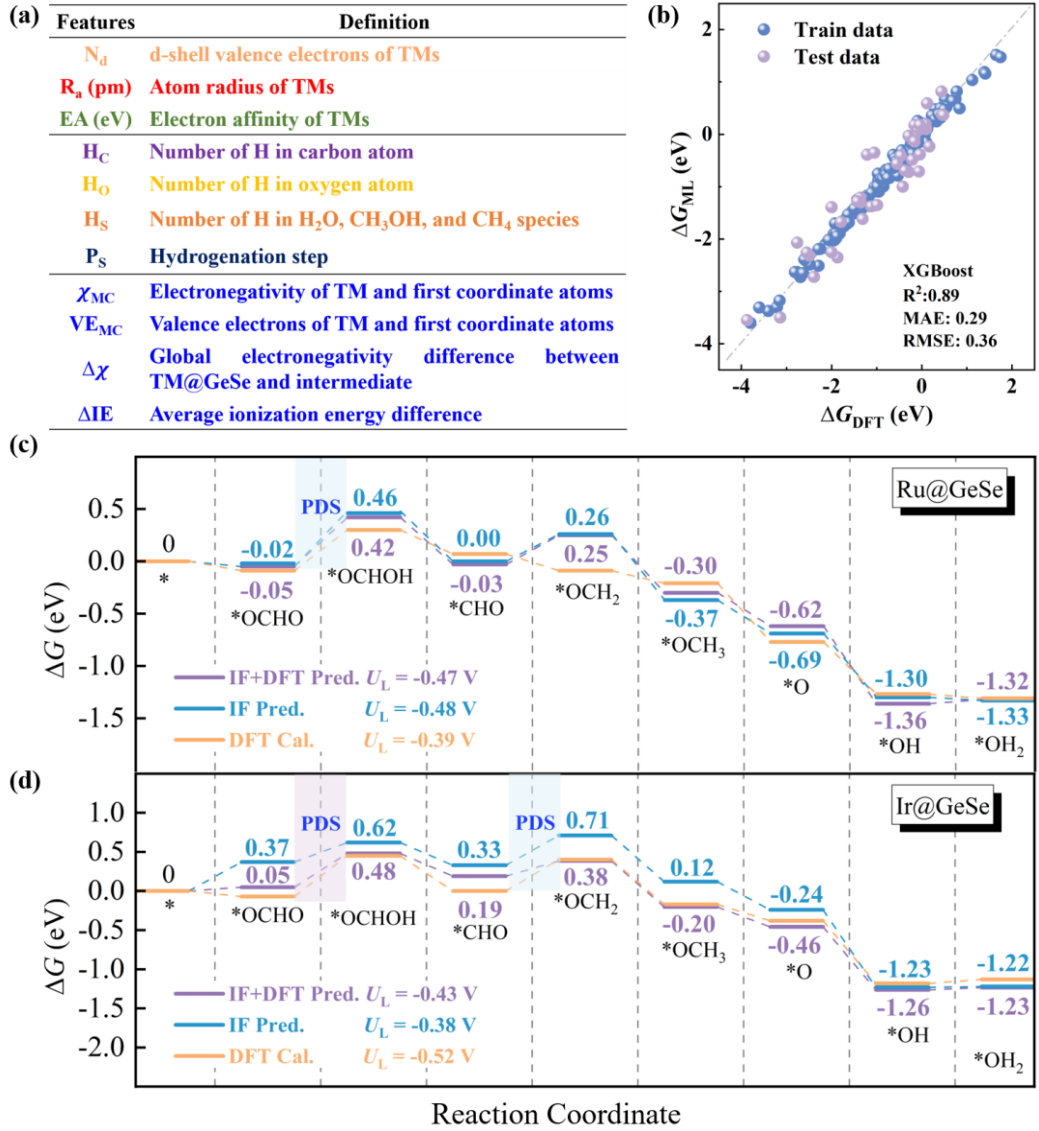

**Figure 6.** (a) ML model with only 11 intrinsic features. (b) Comparison of Δ$G$ values from ML predictions using the XGBoost model with DFT calculations. Free energy diagrams for (c) Ru@GeSe and (d) Ir@GeSe. Yellow, purple, and blue lines represent DFT calculations, ML predictions with IF+DFT features, and ML predictions with IF features, respectively.

$$\chi_{MC} = \chi_M + 3\chi_{Coor} \qquad (4)$$

$$VE_{MC} = VE_M + 3VE_{Coor} \qquad (5)$$

where $\chi_M$ and $VE_M$ represent the electronegativity and the number of valance electrons of the metal atom, respectively. $\chi_{Coor}$ and $VE_{Coor}$ denote the electronegativity and the number of valance electrons of the first coordinated atoms, respectively.

$$\Delta\chi = \overline{\chi_{cat}} - \overline{\chi_{Int}} = \frac{\chi_M + \frac{\sum_{i=1}^{N_{atom}} N_i \chi_i}{N_i}}{2} - \frac{\sum_{j=1}^{N_{atom}} N_j \chi_j}{N_j} \quad (6)$$

where $\overline{\chi_{cat}}$ and $\overline{\chi_{Int}}$ are the average electronegativity of the catalyst and the intermediate, respectively. $\chi_i$ implies the electronegativity of each atom in catalyst except the metal atom, and $\chi_j$ refers to the electronegativity of each atom in the intermediate.

$$\Delta IE = \frac{1}{C}\sum_{i=1}^{C} IE_M - IE_i \quad (7)$$

where $IE_M$ and $IE_i$ represent the ionization energy of the metal atom and its coordination atoms (C or O) in the intermediate, respectively.

The XGBoost model utilizing only intrinsic features also performed well ($R^2 = 0.89$, MAE = 0.29 eV, and RMSE = 0.36 eV), as shown in Figure 6(b). The ML predictions matched well with DFT calculations (Figure S17). Features importance analysis highlighted $N_d$ as the most significant descriptor (Figure S18). To exemplify the agreement between predicted and DFT-calculated $\Delta G$ values, we used Ru@GeSe as an example. In the $CO_2RR$ process, the step from *OCHO to *OCHOH was the PDS for both the XGBoost model with IF+DFT and IF features, and the DFT calculations. Specifically, the $\Delta G_{max}$ was predicted to be 0.47 eV and 0.48 eV by the XGBoost model with IF+DFT and IF features, respectively, compared to 0.39 eV from DFT calculation, indicating good agreement with the DFT results, as seen in Figure 6(c).

To validate the model's reliability, we used Ir@GeSe as a test case. As illustrated in Figure 6(d), the ML predictions using IF+DFT features matched the DFT results better than those only IF features. Specifically, the step from *OCHO to *OCHOH was the PDS for ML predictions using IF+DFT features and DFT calculations, with $\Delta G_{max}$ values of 0.43 eV and 0.52 eV, respectively. But the PDS for ML predictions using IF features was the step from *CHO to *OCH$_2$, with a $\Delta G_{max}$ of 0.38 eV.

To quickly predict the final product, we introduced producing probability ($P_n$) to differentiate HCOOH, $CH_4$, and $CH_3OH$. As depicted in Figure 7(a), $P_1$ and $P_2$ were defined as selectivity descriptors to differentiate HCOOH from $CH_3OH/CH_4$, and $CH_3OH$ from $CH_4$. For instance, when $\Delta\Delta G_1 > 0.2$ eV, $P_1$ is between 0 and 0.45, favoring HCOOH. When $P_1$ lies between 0.55 and 1.0, $CH_4/CH_3OH$ is more likely. Similarly, when $\Delta\Delta G_2 > 0.2$ eV, $P_2$ falls within the range of 0 to 0.45, favoring CH$_3$OH. When $\Delta\Delta G_2 < -0.2$ eV, $P_2$ is in the range of 0.55~1.00, favoring CH$_4$. Otherwise, both CH$_4$ and CH$_3$OH are possible.

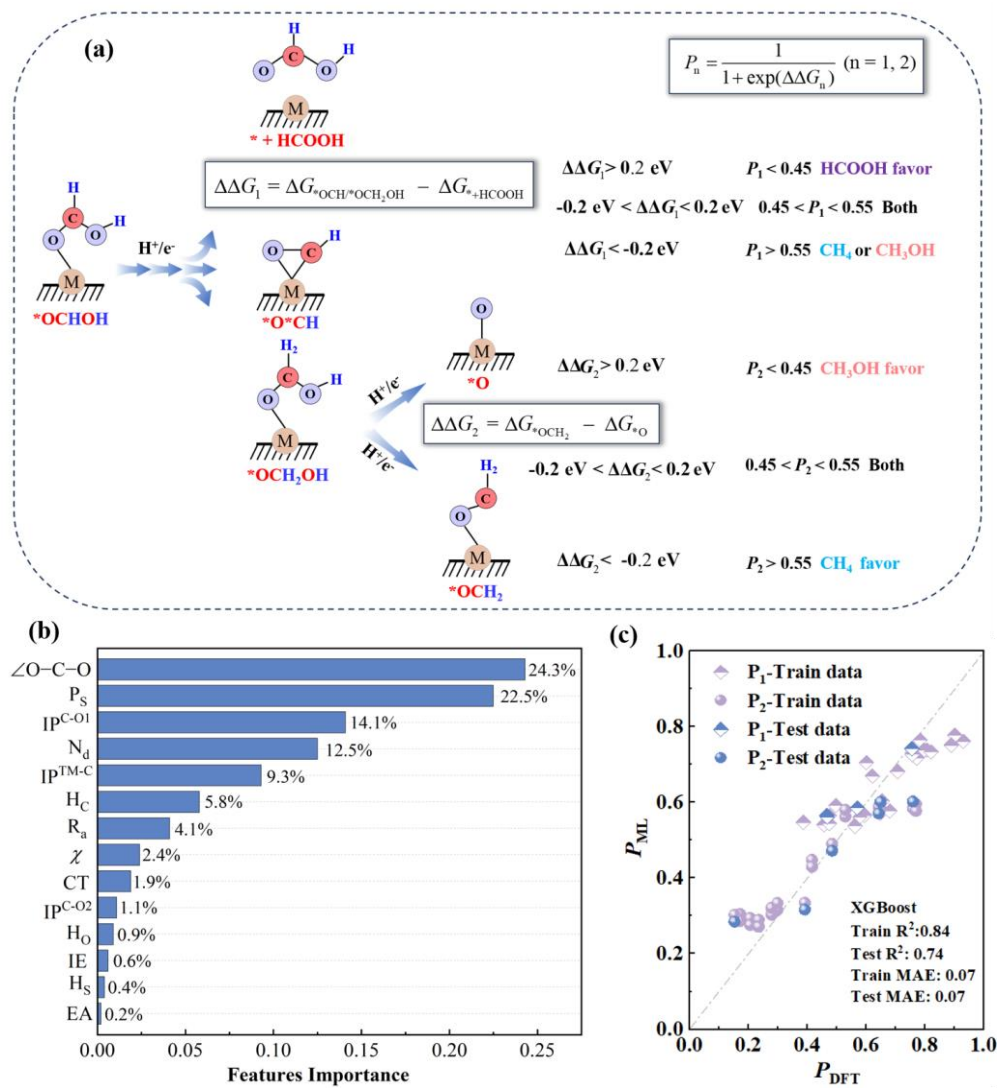

**Figure 7.** (a) Definition of product probability ($P_n$) to differentiate HCOOH, CH$_3$OH, and CH$_4$ products. (b) Features importance assessment in XGBoost model with IF+DFT features in product predictions. (c) Comparison of $P$ values from ML predictions using the XGBR model with DFT calculations.

Next, ML models were constructed to predict the formation of products using two schemes: 14-features (IF+DFT) and 11-features (IF). The XGBoost model using 14-features made good predictions for product probabilities, with $R^2 = 0.84$ and MAE = 0.07 eV for training data, and $R^2 = 0.74$ and MAE = 0.07 eV for test data (Figure 7c). Due to the complexity of CO$_2$RR, the ML models still face challenges in predicting product formation. Features importance analysis revealed that ∠O−C−O, $P_S$, and IP$^{C-O1}$ were the most important, contributing 24.3%, 22.5%, and 14.1%, respectively, underscoring the importance of CO$_2$ activation in predicting products.

## 4. Conclusion

Using DFT calculations, we systematically investigated the stability, $CO_2$ activation, selectivity, and activity of TM@GeSe catalysts, identifying Ni, Ru, and Rh@GeSe as highly promising candidates for $CO_2RR$. By utilizing descriptor analysis with 14 features, the XGBoost model achieved accurate predictions of $\Delta G$ values ($R^2$ = 0.92, MAE = 0.24 eV) and product selectivity ($R^2$ = 0.74, MAE = 0.07 eV). The ∠O−C−O and $IP^{C-O1}$ features emerged as crucial for both predictions, highlighting the importance of $CO_2$ activation. By adopting non-DFT-based features, the XGBoost model maintained strong predictive accuracy ($R^2$ = 0.89, MAE = 0.29 eV) for rapid screening. These results not only highlight exceptional SACs for $CO_2RR$ but also provide a scalable and efficient strategy for catalyst design.

## Declaration of Competing Interest

The authors declare that they have no known competing financial interests or personal relationships that could have appeared to influence the work reported in this paper.

## Data availability

The authors declare that the data will be made available on request.

## Acknowledgements

This work is supported by the Natural Science Foundation of China (Grants No. 12374061) and the KC Wong Magna Foundation at Ningbo University. The computations are supported by the High-Performance Computing Center of Ningbo University.

## Appendix A. Supplementary material

Supplementary data associated with this article can be found in the online version at doi: